\documentclass[12pt]{article}
\usepackage{amssymb,amsmath,latexsym}
\usepackage{graphicx}
\usepackage[final]{pdfpages}
\usepackage{verbatim}
\usepackage{color}
\usepackage{amsfonts,amsthm,hyperref,mathrsfs,ulem,tikz}

\usepackage{cancel} 
\usepackage{soul}

\usetikzlibrary{arrows,patterns}

\title{Levinson theorem for discrete Schr\"odinger operators \\ on the line with matrix potentials having a first moment}

\author{Miguel Ballesteros$^1$, Gerardo Franco~C\'ordova $^1$, \\ Ivan Naumkin$^1$,   Hermann Schulz-Baldes$^2$
	\\
	\\
	{\small $^1$IIMAS, UNAM, Mexico}
	\\   
	{\small $^2$ Friedrich-Alexander-Universit\"at Erlangen-N\"urnberg, Department Mathematik, Germany}
}

\date{ }

\newtheorem{theo}{Theorem}
\newtheorem{defini}[theo]{Definition}
\newtheorem{proposi}[theo]{Proposition}
\newtheorem{lemma}[theo]{Lemma}

\newtheorem{coro}[theo]{Corollary}
\newtheorem{rem}[theo]{Remark}

\newcommand{\CM}{{\mathbb C}}
\newcommand{\NM}{{\mathbb N}}
\newcommand{\RM}{{\mathbb R}}
\newcommand{\SM}{{\mathbb S}}

\newcommand{\ZM}{{\mathbb Z}}

\newcommand{\DM}{{\mathbb D}}

\newcommand{\Ss}{{\cal S}}
\newcommand{\Oo}{{\cal O}}
\newcommand{\Tr}{\mbox{\rm Tr}}

\newcommand{\Mm}{{\cal M}}

\newcommand{\Ran}{\mbox{\rm Ran}}
\newcommand{\Ker}{\mbox{\rm Ker}}

\newcommand{\one}{{\bf 1}}

\newcommand{\overz}{1/z}

\newcommand{\bsm}{\left(\begin{smallmatrix}} 
\newcommand{\esm}{\end{smallmatrix}\right)}  
\definecolor{GR}{rgb}{.35,.7,.35}

\begin{document}

\maketitle

\begin{abstract}
This paper proves new results on spectral and scattering theory for matrix-valued Schr\"odinger operators on the discrete line with non-compactly supported perturbations whose first moments are assumed to exist. In particular, a Levinson theorem is proved, in which a relation between scattering data  and spectral properties (bound and half bound states)   of the corresponding Hamiltonians is derived. The proof is based on stationary scattering theory with prominent use of Jost solutions at complex energies that are controlled by Volterra-type integral equations.	
\end{abstract}

\vspace{.5cm}


\section{Introduction}

This paper proves a Levinson theorem for a matrix-valued Schr\"odinger operator of the form $H=H_0+V$ on the Hilbert space $\ell^2(\mathbb{Z},\mathbb{C}^L)$ with $L\in\NM$, where the free operator $H_0$ is the discrete Laplacian, up to an additive constant, given by
\begin{equation}\label{IntE1}
H_0 u (n) := u(n+1) + u(n-1),    \qquad u \in \ell^2(\mathbb{Z},\mathbb{C}^L), 	 
\end{equation} 
and $V$ is a potential energy given by a self-adjoint matrix-valued multiplication operator 
$$ 
 V u (n) = V(n) u(n),
$$
where $V(n) \in \mathbb{C}^{L\times L} $ is a self-adjoint $L\times L$ matrix for every $n \in \mathbb{N}$. The main assumption is that its first moment exists, {\it i.e.}
\begin{equation}\label{IntE2}
	\sum_n   \|  n V(n) \| < \infty. 
\end{equation}
Levinson theorem, see Theorem~\ref{theo-Levinson} below,  relates scattering data to the number of bound and half-bound states  of $H$. The objects of scattering theory are formulated in terms of Jost solutions which are formal eigenvectors of $H$ with prescribed asymptotic behavior. More precisely,  we will identify $ H$, $H_0$ and $V$ with extended operators acting on sequences $u\in  ( \mathbb{C}^L)^\mathbb{Z} $ or  $u\in\Mm^\mathbb{Z}$ where $\Mm= \CM^{L \times L}$, given by the same formulas, and then consider the formal (or generalized) eigenvalue equation
\begin{align}\label{IntE4}
	H u = E u, 
\end{align}	
where $  E \in \mathbb{C} $ is parameterized in the following form 
\begin{equation}\label{IntE6}
E = z+ 1/z,   \qquad     z \in \mathbb{C}.
\end{equation}
Then the Jost solutions $u^z_{\pm}\in  \Mm^\mathbb{Z}$ are
specified by
\begin{align}\label{IntE7}
	u^z_{\pm}(n)=z^{n}(\mathbf{1}+o(1)) , \qquad n \to \pm \infty, 
\end{align}
where $\mathbf{1}$ is the identity matrix in $\Mm$. As $n\mapsto z^{n}\mathbf{1}$ are the solutions of the free equation $H_0u=Eu$, one can alternatively state that the Jost solutions $u^z_{\pm}$ are those solutions of \eqref{IntE4} that behave as the free solutions asymptotically at $\pm\infty$. In Section \ref{sol-fun}, we prove their existence and several of their properties. Our previous work \cite{BFGS} only addresses the case $0 < |z| \leq 1$, which is hence here extended to cover $ |z |> 1 $. The full picture, that is $ z \in \mathbb{C}\setminus \{ 0 \}  $, and furthermore a discussion of the case $z = 0$ are necessary for a proof of Levinson's theorem. The following elementary remark allows to introduce the scattering matrix and will be referred to at several reprises.

\begin{rem}\label{fun_sol}
{\rm For each $z\in \mathbb{C}\setminus\{1,0,-1\}$, due to the asymptotic behavior \eqref{IntE7} of {the Jost} solutions, the columns of the matrix $( u^{z}_{\pm}, u^{1/z}_{\pm} )$ are linearly independent for $ z \in  \mathbb{C} \setminus  \{0,-1, 1 \} $ and, therefore,  they form a basis of solutions of the generalized eigenvalue equation \eqref{IntE4}}. 
\hfill $\diamond$
\end{rem} 

\begin{defini}\label{dis_coe}
  For $  z\in \mathbb{C}\setminus\{1,0,-1\}$, we denote by  $M_{\pm}^z$ and $N_{\pm}^z$ the $L\times L$ matrices satisfying 
  \begin{equation}\label{lineal}
    u^z_{+}=u^{z}_{-}M_+^z+u^{1/z}_{-}N_+^z, \qquad  u^{1/z}_{-}=u^{1/z}_{+}M_-^{z}+u^{z}_{+} N_-^{z}.
  \end{equation}
\end{defini}

The matrices  $M_{\pm}^z$ and $N_{\pm}^z$ are the key ingredients in stationary scattering  theory.  The scattering matrix is built from them in the following way:  
 \begin{defini}\label{disper}
 For $z\in\mathbb{C}  \setminus\{1,0,-1\}   $ with $|z|\leq1$ such that $M_{\pm}^z$ is invertible, the scattering matrix is given by
 \begin{equation}
 \label{eq-ScatMat}
 \mathcal{S}^z=\begin{pmatrix}
 	(M^z_+)^{-1} & -N_-^z(M_-^z)^{-1}\\
 	-N_+^z(M_+^z)^{-1} & (M_-^z)^{-1}
 	\end{pmatrix} .
\end{equation}
 The entries of $\Ss^z$ also define the transmission and reflection coefficients matrices by
 $$
 \mathcal{S}^z=\begin{pmatrix} T^z_+ & R^z_- \\ R^z_+ & T^z_- \end{pmatrix} .
$$
 \end{defini}
 
Proposition \ref{invertible} below implies that $ \Ss^z $ is well-defined and unitary for $z \in \mathbb{S}^1 \setminus \{1,-1 \}$. 
Moreover, Proposition~\ref{Men0} combined with Proposition~\ref{split}  implies that the function $z\in\mathbb{S}^{1}\setminus \{-1,1 \}  \mapsto \det(\mathcal{S}^z) $ is differentiable. Proposition \ref{anal_1} and Remark \ref{para_-} below show that $M^z_{\pm}$ are invertible in a neighborhood of $1$ in $\overline{\mathbb{D}} \setminus \{-1, 1 \}$, and similarly in a neighborhood of $ -1$.	

\vspace{.2cm}

As a final preparation for the statement of the main result, let us introduce the path $\Gamma_+^\epsilon$ for  $\epsilon>0$ as the truncated upper semicircle parameterized by $\gamma_+^\epsilon:[0,1] \to \mathbb{S}^1$ given by  
$$
\gamma_+^\epsilon(t)=e^{\imath\pi((1-t)\epsilon+t(1-\epsilon))}.  
$$

\begin{theo}[Levinson Theorem]
\label{theo-Levinson}
The Hamiltonian $H$ has only a finite number $J_b$ of eigenvalues  $E_1,\ldots,E_{J_b}\in\RM$  {\rm (}listed with their multiplicity{\rm )} and they are outside of $[-2,2]$. Moreover, at the thresholds $E=\pm 2$, there are $J^\pm_{h}\leq L$ linearly independent bounded solutions of $Hu=\pm 2 u$ which are called half-bound states. With $J_h=J_h^-+J^+_h$, one has
	$$
	2\pi \imath(J_b\,+\,\tfrac{1}{2}\,J_h\,-\,L) 
	\;=\;-\,\lim\limits_{\epsilon \to 0}\int_{\Gamma_+^\epsilon}\det(\mathcal{S}^z)^{-1}\frac{d}{dz}\det(\mathcal{S}^z)dz
	\;.
	$$ 
\end{theo}

This article  does not assume that the potential is compactly supported, as we did in our previous work  \cite{BFS}.  Non assuming compactly supported potentials requires different techniques: the compactly supported case \cite{BFS} is prominently based on
transfer matrices. The existence and differentiability of the Jost solutions in \cite{BFS}  pends on transfer matrix techniques that, to our best knowledge, do not transpose to the case with non-compact support. This implies that the proofs in this manuscript differ substantially from  \cite{BFS}. As already stated above, the Jost solutions are here studied as solutions of integral equations, similarly as in  \cite{BFGS}.  To avoid overlaps with the earlier works, we state needed results from \cite{BFS, BFGS}  without proofs and focus on  the innovative aspects of the arguments. 

\vspace{.2cm}

Finally, let us give a brief account of earlier related works on scattering for one-dimensional discrete Schr\"odinger operators.  Foundations and inverse scattering theory for the scalar case are laid out in \cite{Cas,Gus,Gus2,Gus3,AN,Tes}. Levinson's theorem for one-dimensional discrete operators in the scalar case is proved in \cite{HKS} and more recently in \cite{CS,IT,NRT}. Scattering in a periodic background is treated in \cite{EMT}. Works on the scattering theory for the matrix-valued case are scarce  \cite{Ser,BAC1,BAC2}, but the latter two also construct Jost solutions and a scattering matrix under a moment condition similar as is done below. What is missing in \cite{BAC2}, however, is the fine analysis of the analytic behavior of the Jost solutions and the scattering matrix at the band edges so that the authors could not conclude that there is a finite number of bound states nor analyze half-bound states nor prove a Levinson theorem. For scattering theory for continuous one-dimensional Schr\"odinger operators with a matrix-valued potential, there is also abundant literature, most of which is cited in the recent monograph by Aktosun and Weder \cite{AW}. A Levinson theorem in that framework is proved in \cite{AW0,ACP}, and an index-theoretic perspective is given in \cite{KR} (in the scalar case, but this readily transposes to the matrix-valued case, {\it e.g.} \cite{BS,IT}). Complementary references for scattering theory on matrix Schr\"odinger operators and inverse scattering can be found in \cite{Mar,AW}.

\vspace{.2cm}

This paper is organized as follows. Section~\ref{sol-fun} constructs Jost solutions and proves some key regularity properties and estimates for them. Section~\ref{SScattering} then deduces analytical properties of the scattering matrix. Sections~\ref{halfbound-states} and \ref{bound-states} address respectively the half-bound and bound states appearing in Levinson's theorem. Section~\ref{STime} gives a formula for the time delay that was already used in \cite{BFS} and for which a new simplified proof is provided here. In Section \ref{Slevinson}, we prove our main result (Levinson Theorem) using the results and constructions of  the other sections.

\section{Jost solutions}\label{sol-fun}

\subsection{Existence}

In this section, we construct  fundamental solutions to  Eq. \eqref{IntE4}.

\begin{defini}\label{FreeSolo1}
For every  $z\in \mathbb{C} \setminus\{0\} $, we denote by $s^z$ the scalar solutions $s^z\in\CM^\ZM$ of $H_0u=(z+1/z)u$ such that  $s^z(0) = 0$, $s^z(1) = 1$.
\end{defini}

Explicitly, one can verify that 
\begin{align}\label{eq-sz}
s^z(n)= \left\{ \begin{array}{cc}
\frac{1}{z - z^{-1}}(z^{n}-z^{-n}) , \;\;\;\;\;&   z^2\neq 1, \\
(\pm1)^{n+1}n,  & z= \pm1 .
\end{array} \right.
\end{align}	

\begin{proposi}[Fundamental solution]\label{jost-sol-1}
  For every $z\in \mathbb{C} \setminus \{ 0 \}$, there exist solutions $u^z_{\pm} \in \Mm^{\mathbb{Z}}$  to Eq. \eqref{IntE4} with $E=z+\frac{1}{z}$, such that
  \begin{align}\label{u-asin}
    u^z_{\pm}(n)=z^{n}(\mathbf{1}+o(1)) , \qquad n \to \pm \infty. 
  \end{align}
  For $0<|z|\leq 1$, they satisfy: 
  \begin{equation}\label{6}
  \begin{aligned}
  u_{+}^z(n)&=z^{n}\mathbf{1} - \sum_{j=n+1}^{\infty} s^z(j-n)V(j)u_{+}^z(j) , \qquad n\in \mathbb{Z},\\
  u_{-}^{1/z}(n)&=z^{-n}\mathbf{1} + \sum_{j=-\infty}^{n-1} s^{1/z}(j-n)V(j)u_{-}^{1/z}(j) , \qquad n\in \mathbb{Z}. 		 
  \end{aligned}	
  \end{equation}
  Moreover, if one defines 
  $$\tilde{u}_{\pm}^z(n):=z^{-n}u_{\pm}^z(n) , \qquad n\in  \mathbb{Z}, \ z\in \mathbb{C}\setminus\{0\},$$   
  
   $$\tilde{u}^0_{+}(n)=\mathbf{1}, \qquad \tilde{u}^{1/z}_{-}(n)\big|_{z=0}=\mathbf{1},$$ then for each $n\in \mathbb{Z}$ the functions $z \mapsto \tilde{u}_{+}^z(n)$ and $z \mapsto \tilde{u}_{-}^{1/z}(n)$ are analytic on the open unit disc
\begin{equation}\label{disc}
		\DM:= \{z\in \mathbb{C}: |z|< 1\},
	\end{equation}
and continuous on its closure $\overline{\DM}=\{z\in \mathbb{C}: |z|\leq 1\}$ with a uniform bound in $n$ and $z\in\overline{\DM}$. 
\end{proposi}
\noindent {\bf Proof.} Existence, analyticity
 and continuity of the solutions
  $u^z_{+}$ and $u^{1/z}_{-}$, for $0<|z|\leq 1$,
   was already
  proved  in Lemma 7 of \cite{BFGS} which also contains \eqref{6}. The argument is essentially based on solutions of the Volterra equation as stated in Theorem~\ref{volterra}.

\vspace{.1cm}

The extension of   $\tilde{u}^z_{+}$ 
   and $\tilde{u}^{1/z}_{-}$ to $ z = 0$
    is derived   using the proof of Lemma 7 
   in \cite{BFGS} setting  $K^0(n,j)=0$.
   We recall that 
   $  K^z(n,j)=-z^{j-n}s^z(j-n)V (j) $ in \cite{BFGS}. 
    Notice   
   that $\lim_{z\to 0}z^m s^z(m) \to 0$ for $m\geq 0$, 
   which implies that $z \mapsto K^z(n,j)$ 
   is analytic on $|z|<1$. 
The uniform bound on $\tilde{u}^z_\pm(n)$, w.r.t. $n$ and $z$,  follows again from Theorem~\ref{volterra}.

\vspace{.1cm}
   
Next we construct the solution $u^z_+$ for $|z|>1$, 
   the construction of the solution $u^{1/z}_-$ 
   is analogous. Let $z\in \mathbb{C}$ with $|z|>1$
    and take $m\in \mathbb{N}$ such 
    that 
\begin{equation}
\label{eq-TailBound}
\sum_{j=m}^{\infty} \|V(j)\| < \left|\frac{z^2-1}{2z}\right| .
\end{equation} 
Consider     the Volterra-type equation for $n>m$ 
\begin{equation}\label{sat}
	Y(n)=z^{n}\mathbf{1}+\frac{z}{z^2-1}\sum_{j=n+1}^{\infty} z^{n-j}V(j)Y(j)+\frac{z}{z^2-1}\sum_{j=m}^{n}z^{j-n}V(j)Y(j),
\end{equation} 
 which is equivalent to Eq. \eqref{IntE4} (this is verified in \eqref{cuenta} below, see also \cite{Agranovich} p. 31 for the continuous setting).
In order to find a solution $Y$ to Eq. \eqref{sat}, let us set $X(n)=z^{-n}Y(n)$ and rewrite  Eq. \eqref{sat} in the next way 
\begin{equation}\label{sum_equ2}
	X(n)=\mathbf{1}+\frac{z}{z^2-1}\sum_{j=n+1}^{\infty}V(j)X(j)+\frac{z}{z^2-1}\sum_{j=m}^{n}z^{2(j-n)}V(j)X(j). 
\end{equation}  
For each $n,j \geq m$, one hence defines $K(n,j) \in \Mm$   by
$$
K(n,j)=\left\{ \begin{array}{cc}
 \frac{z}{z^2-1} z^{2(j-n)}V(j), &   m\leq j \leq n, \\
 \frac{z}{z^2-1}V(j), &  n+1 \leq j.
 \end{array} \right.$$   
The definition implies that  $\|K(n,j)\| \leq \left|\frac{z}{z^2-1} \right| \|V(j)\|,$ for all $n,j \geq m$. Next consider the operator $T:\ell^{\infty}([m,\infty)\cap \mathbb{N},\Mm) \to  \ell^{\infty}([m,\infty)\cap \mathbb{N},\Mm)$ defined by $$(TX)(n)= \sum_{j=m}^{\infty}K(n,j)X(j).$$
  Notice that $T$ is well-defined because one has   due to \eqref{eq-TailBound}
$$\sum_{j=m}^{\infty}\|K(n,j)X(j)\|\leq \|X\|_{\infty}\sum_{j=m}^{\infty}\|K(n,j)\|\leq \|X\|_{\infty}\left|\frac{z}{z^2-1} \right| \sum_{j=m}^{\infty}\|V(j)\|< \frac{1}{2} \|X\|_{\infty}
$$  
for all $n \in [m,\infty) \cap \mathbb{N}$, 
where $\| \cdot \|_{\infty}$ denotes the norm in the Banach space $\ell^{\infty}([m,\infty)\cap \mathbb{N},\Mm)$. It follows that the operator norm satisfies
 $$\|T \| <\frac{1}{2} .$$
 Eq. \eqref{sum_equ2} takes the form of  
 $$X=\mathbf{1}+TX.$$  Therefore its solution is 
$$
X = (1 - T)^{-1}\mathbf{1}\in\ell^\infty(\NM,\Mm), 
 $$
which exists because $ \| T\| \leq \frac{1}{2} .$ 
 It is easy to see from Eq. \eqref{sum_equ2} that $X(n) \to \mathbf{1}$ for $ n \to +\infty.$ Now we define for $n\geq m$, 
 $$
 u_+^z(n) := Y(n),
 $$
  and recursively for $n\leq m$, $$u_+^z(n-1):=Eu_+^z(n)-V(n)u_+^z(n)-u_+^z(n+1).$$
    It follows from its definition that $u_+^z$ satisfies \eqref{u-asin}, and for $n\leq m$ it satisfies \eqref{IntE4}. The fact that $u_+^z$ satisfies \eqref{IntE4} for $n\geq m+1$ follows from its definition and the next computation:
 \begin{align}
 \label{cuenta}
 	Y&(n+1)+Y(n-1)
\nonumber
\\
&
= (z^{n+1}+z^{n-1})\mathbf{1}+\frac{z}{z^2-1}\left(\sum_{j=n+2}^{\infty} z^{n+1-j}V(j)Y(j)+\sum_{j=n}^{\infty} z^{n-1-j}V(j)Y(j)\right)\nonumber
 	\\&\;\;\; \;\;+\frac{z}{z^2-1}\left(\sum_{j=m}^{n+1}z^{j-n-1}V(j)Y(j)+\sum_{j=m}^{n-1}z^{j-n+1}V(j)Y(j)\right)\nonumber
 	\\&= \notag Ez^n\mathbf{1}+\frac{z}{z^2-1}\left(\sum_{j=n+1}^{\infty} (z^{n+1-j}+z^{n-1-j})V(j)Y(j)\right)
 	\\ \notag & \;\;\; \;\;+\frac{z}{z^2-1}\left(\sum_{j=m}^{n}(z^{j-n-1}+z^{j-n+1})V(j)Y(j)\right)\nonumber\
 	\\ \notag &\;\;\; \;\; +\frac{z}{z^2-1}(-V(n+1)Y(n+1)+z^{-1}V(n)Y(n)+V(n+1)Y(n+1)-zV(n)Y(n))\nonumber\\&=Ez^n\mathbf{1}+E\frac{z}{z^2-1}\sum_{j=n+1}^{\infty} z^{n-j}V(j)Y(j)+E\frac{z}{z^2-1}\sum_{j=m}^{n}z^{j-n}V(j)Y(j)-V(n)Y(n)\nonumber\\&=EY(n)-V(n)Y(n).
 \end{align}
 This completes the proof.
\hfill $\Box$

\subsection{Derivatives of Jost solutions} \label{sderivatives}

The previous section constructed the solutions $u^z_+$ and $u^{1/z}_-$ and demonstrated their analyticity on open unit disc $\DM$ 
as well as their continuity on its closure $\overline{\DM}$.  In this section, it is proved they are continuously  differentiable   $\overline \DM$ in the sense of the definition below. The proof is inspired by the continuous case that is presented in Deift and Trubowitz \cite{Dei}.

\begin{defini}
Suppose that $U$ is a subset of $\mathbb{C}$.  
A function $g: U \to {\mathbb{C}} $ is said to be differentiable at $z \in U$ with differential (or derivative) $ \frac{d}{dz} g (z)$ or $  \dot{g}(z) $ if for every $\epsilon > 0$ there exists $\delta > 0$ such that  
$$
0 < |h| < \delta, \: z + h \in U\;\; \Longrightarrow   \;\;
\Big | \frac{ g(z+h) - g(z) }{h}  - \dot g (z)\Big | < \epsilon.  
$$
Then $g$ is said to be differentiable on $U$ if it is differentiable at every point of $U$, and continuously differentiable on $U$ if its derivative is a continuous function. Likewise, these concepts are defined for vector or matrix values functions. 
\end{defini}

Notice that if  $z$ is an interior point of $U$, this coincides with the usual definition of analyticity.  For the proof of the differentiability of $u^z_+$, we will again use the sequence $\tilde{u}_{+}^z(n)=z^{-n}u_{+}^z(n)$. Due to \eqref{6}, it satisfies
$$
\tilde u_{+}^z(n)=\mathbf{1} + \sum_{j=n+1}^{\infty} z^{j-n}s^z(j-n)V(j)\tilde u_{+}^z(j),
\qquad
z\in\overline{\DM}.
$$
Hence let us set, for $z \in \overline \DM$ and $j> n \in \mathbb{N}$,
$$
H(z,n,j)=-z^{j-n}s^z(j-n)=-\sum_{m=0}^{j-n-1} z^{2m+1},
$$ 
so that
\begin{equation}\label{sati}
\tilde u_{+}^z(n)=\mathbf{1} + \sum_{j=n+1}^{\infty} H(z,n,j)V(j)\tilde u_{+}^z(j).
\end{equation} 
From the definition, one readily checks that
\begin{equation}\label{coth}
	|H(z,n,j)|\leq j-n,
\end{equation} and $$\dot H(z,n,j)=\frac{2(j-n)z^{j-n}}{1-z^2}+\frac{z^{2(j-n)}-1}{1-z^2} \cdot\frac{1+z^2}{1-z^2}= \frac{2(j-n)z^{j-n}}{1-z^2}-\frac{1+z^2}{1-z^2}\sum_{m=0}^{j-n-1}z^{2m}. $$
It follows that
\begin{equation}\label{cothde}
	|(1-z^2)\dot H(z,n,j)|\leq 4(j-n).
\end{equation}
Now formally deriving \eqref{sati} w.r.t. $z$ one obtains the  equation 
\begin{equation}
\label{eq-FormalDeriv}
\frac{d}{dz}\tilde u_+^z(n) = \sum_{j=n+1}^\infty \dot{H}(z,n,j)V(j)\tilde u_+^z(j)+\sum_{j=n+1}^\infty H(z,n,j)V(j) \frac{d}{dz}\tilde u_+^z(j), 
\end{equation}
for $n \in \mathbb{N}$. Let us first verify that this equation, multiplied with $1-z^2$, has a bounded solution.

\begin{lemma}\label{derivative}
For $z\in \overline \DM$ there exists a solution $h^z \in \ell^\infty(\NM,\Mm)$ of the integral equation
$$
h^z(n) = \sum_{j=n+1}^\infty (1-z^2)\dot{H}(z,n,j)V(j)\tilde u_+^z(j)+\sum_{j=n+1}^\infty H(z,n,j)V(j) h^z(j),
$$ 
which is analytic on $\DM$ and continuous on $\overline{\DM}$. 
\end{lemma}
 \noindent {\bf Proof.} 
By Proposition~\ref{jost-sol-1}, $\tilde u_+^z\in \ell^\infty(\mathbb{N}, 
\Mm)$ with $ \|  \tilde u_+^z  \|_\infty $ uniformly bounded for $z \in \overline{\DM}$.
For $n \in \mathbb{N}$,  \eqref{cothde} implies that 
	$$\sum_{j=n+1}^{\infty} \| (1-z^2)\dot{H}(z,n,j)V(j)\tilde u_+^z(j)\| \leq \sum_{j=n+1}^{\infty} 4j\|V(j)\|\|\tilde u_+^z(j)\|<4\|\tilde u_+^z\|_\infty \sum_{j=0}^{\infty} j\|V(j)\|.$$ 
This implies that the function \begin{align}\label{nomames}g^z(n):= \sum_{j=n+1}^\infty (1-z^2)\dot{H}(z,n,j)V(j)\tilde u_+^z(j)\end{align} belongs to $\ell^\infty(\mathbb{N}, \Mm)$, is analytic in $z\in\DM$ and uniformly bounded on $\overline{\DM}$:
$$
\|g^z\|_\infty \leq 4\sup_{z\in \overline \DM}\|\tilde u_+^z\|_\infty \sum_{j=0}^{\infty} j\|V(j)\|.
$$ 
Then the result follows from the Theorem \ref{volterra} with $g^z$ defined as above and $K^z(n,j)=H(z,n,j)V(j)$, $M(j)=j\|V(j)\|$.
 \hfill $\Box$ 

\begin{proposi} \label{derdtu}
For each $n\in \mathbb{N}$, the function $z \mapsto \tilde u_+^z(n)$ is continuously differentiable  on $\overline \DM \setminus\{1,-1\}$. Moreover, the derivative $\frac{d}{dz}\tilde u_+^z \in \ell^\infty(\mathbb{N},\Mm)$ and it satisfies \eqref{eq-FormalDeriv}.
\end{proposi}

 \noindent {\bf Proof.} 
For $z\in \overline \DM \setminus\{1,-1\}$, let us set $f^z:=(1-z^2)^{-1} h^z$ where $ h^z $ is defined in  Lemma~\ref{derivative}. Lemma~\ref{derivative} implies that $z\mapsto f^z(n)$ is continuous on $\overline \DM \setminus\{1,-1\}$ and satisfies the following equation
\begin{equation}\label{equf}
	f^z(n)=\sum_{j=n+1}^\infty \dot{H}(z,n,j)V(j)\tilde u_+^z(j)+\sum_{j=n+1}^\infty H(z,n,j)V(j) f^z(j), \qquad  n \in \mathbb{N}.
\end{equation}
Take $n\in \mathbb{N}$ and $z\in \overline \DM \setminus \{1,-1\}$. We prove that  $\frac{d}{dz}\tilde u_+^z(n) = f^z(n)$, then the result follows from the above. Using \eqref{sati} and \eqref{equf} one has 
$$
\frac{\tilde u_+^{z+h}(n) - \tilde u_+^{z}(n)}{h} - f^z(n)= G(h)+ \sum_{j=n+1}^{\infty} H(z,n,j)V(j) \left(\frac{\tilde u_+^{z+h}(j)-\tilde u_+^z(j)}{h}-f^z(j)\right),
$$
where 
$$
G(h)=\sum_{j=n+1}^\infty \left(\frac{H(z+h,n,j)-H(z,n,j)}{h}V(j)\tilde u_+^{z+h}(j)-\dot H (z,n,j)V(j)\tilde u_+^z(j)\right).
$$
Eq. \eqref{cothde} implies  that 
$$\|\dot H(z,n,j)V(j)\tilde u_+^z(j) \|\leq 4 |(1-z^2)^{-1}|j\|V(j)\|,
$$ 
which is hence summable in $j$ by the main assumption \eqref{IntE2}. On the other hand, the estimate 
$$ 
\left \|\frac{H(z+h,n,j)-H(z,n,j)}{h}\right \| \leq \int_{0}^1\|\dot H(z + th,n,j)\| dt 
$$
leads to
$$ 
\left \|\frac{H(z+h,n,j)-H(z,n,j)}{h} V(j) \tilde u_+^{z+h}(j)\right \| \leq 4 \Big ( \int_{0}^1  |1-(z  + th)^2|^{-1} d t \Big ) j  \|V(j)\| ,
$$
which is thus also summable in $j$. Therefore, the Lebesgue dominated convergence theorem implies that $G(h) \to 0$ as $ h \to 0.$ Now by the Gronwall lemma (Lemma \ref{gronwall}) and Eq. \eqref{coth}, one has
$$\left |\frac{\tilde u_+^{z+h}(n) - \tilde u_+^{z}(n)}{h} - f^z(n)\right |\leq |G(h)|\exp\left(\sum_{j=n+1}^{\infty}j \|V(j)\|\right) \;\to\; 0 , \qquad \mbox{as } h \to 0.$$
This implies the desired result.
 \hfill $\Box$ 

\begin{rem}\label{deru}
{\rm Recall that (see Proposition \ref{jost-sol-1}) by definition $u^z_+(n)=z^n \tilde u_+^z(n),$ so that Proposition \ref{derdtu} implies that the map $z \mapsto u^z_+(n)$ is continuously differentiable on $\overline \DM \setminus \{1,-1\}$ for $n\in \mathbb{N}$. Moreover, Eq. \eqref{IntE4} implies that $$u_+^z(n-1)=(z+1/z-V(n))u^z_+(n)-u^z_+(n+1),$$ which along with the above allows to prove that $z \mapsto  u_+^z(n)$ is continuously differentiable on $\overline \DM\setminus\{1,-1\}$ for all $n\in \mathbb{Z}$.	 In a similar way one proves that the map $z\mapsto u^{1/z}_-(n)$ is continuously differentiable on $\overline \DM \setminus\{-1, 0, 1\}$ for all $n\in \mathbb{Z}$. The above results imply that  the map $z \mapsto u^{1/z}_+(n)$ is differentiable for $ z \in (\mathbb{C} \setminus \DM) \setminus \{ -1, 1 \} $, and therefore it is differentiable on $ \mathbb{S}^1 \setminus \{ -1, 1 \} $. The same holds true for the function   $z \mapsto u^{z}_-(n)$.  
} \hfill $\diamond$
\end{rem}

\section{Scattering matrix} 
\label{SScattering}

All formulas in this section are identical to those in \cite{BFS}, but their existence in the present more general context depends on the results of the previous sections. Remark \ref{fun_sol} allows us to define the scattering coefficients $M_{\pm}^z$ and $N_{\pm}^z$  for $z\in \mathbb{C}\setminus\{1,0,-1\}$ (see Definition \ref{dis_coe}). 
These matrices have representations in terms of the Wronskian which for two functions $u,v\in\Mm^\ZM$ is defined by
\begin{equation}\label{wronskian}
	W(u,v)(n)=\imath(u(n+1)^*v(n)-u(n)^*v(n+1)).
\end{equation}
Using the eigenvalue Eq. \eqref{IntE4}, an elementary calculation implies that  $ W(u^{\overline{z}}_{+},u^{z}_{+})(n) $ and  $ W(u^{1/\overline{z}}_{\pm},u^{z}_{\pm})(n)  $ do not depend on $n$ (for every $z$). Then, using the asymptotic behavior 
of Jost solutions, one concludes that for $0<|z|\leq 1$ 
\begin{align}\label{w-asin-1}
	W(u^{\overline{z}}_{+},u^{z}_{+})=0 =	W(u^{1/\overline{z}}_{-},u^{1/z}_{-}),
\end{align}
and for $z\in \mathbb{C}\setminus\{0\}$
\begin{align}\label{w-asin-2}
	W(u^{1/\overline{z}}_{\pm},u^{z}_{\pm})=(\nu^z)^{-1}\mathbf{1}.
\end{align}
where we set
\begin{equation}\label{lanu}
	\nu^z=\frac{\imath}{z-z^{-1}}.
\end{equation}

\begin{proposi}\label{ident}
  For every  $0<|z|\leq 1$ with $z^2\neq 1 $, the following expressions hold true:
  \begin{equation}\label{M_+}
  	M_{+}^z=\nu^z \ W(u_-^{1/\overline{z}},u_+^z),
  	\end{equation}
  	\begin{equation}\label{M_-}
  	M_{-}^z=-\nu^z \ W(u_+^{\overline{z}},u_-^{1/z}).   
  	\end{equation}
  Moreover,  for every  $ |z|\geq 1$ with $z^2\neq 1,$
  	\begin{equation}\label{N_+}
  	N_{+}^z=-\nu^z \ W(u_-^{\overline{z}},u_+^z),  
  	\end{equation}
  	\begin{equation}\label{N_-}
  	N_{-}^z=\nu^z \ W(u_+^{1/\overline{z}},u_-^{1/z}).
  	\end{equation}
\end{proposi}

 \noindent {\bf Proof.} 
Let us start computing $W(u_-^{1/\overline{z}},u_+^z)$ for $0<|z|\leq1$ using equations \eqref{lineal}, \eqref{w-asin-1} and \eqref{w-asin-2}:
\begin{align*}
	W(u_-^{1/\overline{z}},u_+^z)=W(u_-^{1/\overline{z}},u^{z}_{-}M_+^z+u^{1/z}_{-}N_+^z)=W(u_-^{1/\overline{z}},u^{z}_{-})M_+^z+ W(u_-^{1/\overline{z}},u^{1/z}_{-})N_+^z=(\nu^z)^{-1}M_+^z.
\end{align*}
In a similar fashion using that $\nu^{1/z}=-\nu^z$ one has:
\begin{align*}
W(u_+^{\overline{z}},u_-^{1/z})=W(u_+^{\overline{z}},u^{1/z}_{+}M_-^{z}+u^{z}_{+} N_-^{z})=W(u_+^{\overline{z}},u^{1/z}_{+})M_-^{z}+W(u_+^{\overline{z}},u^{z}_{+} )N_-^{z}=-(\nu^z)^{-1}M_-^z.
\end{align*}
The rest of the proof is derived in a similar way.
 \hfill $\Box$ 

 \vspace{.2cm}
 
Proposition \ref{ident} implies 
\begin{equation}\label{masamenos}
(M_+^z)^*=M_-^{\overline{z}}\;\;\mbox{ for }z\in \overline{\mathbb{D}}\setminus\{0\}, \qquad 	(N_+^z)^*=-N_-^{z}\;\;\mbox{ for } \ z\in \mathbb{S}\setminus\{1,-1\}. 
\end{equation}
Next further properties of these  coefficients are proved.
\begin{lemma}
	For every $ z \in  \mathbb{S}^1 \setminus  \{-1, 1 \} $, the following identities hold true: 
	\begin{align}
	\label{I2}
	(M^z_-)^*M^z_-\;&=\;\one\,+\,(N^z_-)^*N^z_-\;,\\\label{I3}
	M^z_+N^z_-\;& =\;-\,N^{\overz}_+M^z_-\;,\\\label{I6}
	(M^z_+)^*M^{z}_+\;&=\;\one\,+\,(N^z_+)^*N^z_+,\\ 
	\label{I5}
	M^{z}_-N^{z}_+\;&=\;-\,N^{\overz}_-M^z_+\;. 
	\end{align}
\end{lemma}

 \noindent {\bf Proof.} 
Equations \eqref{lineal} and \eqref{w-asin-2} imply that
\begin{align}
\label{to2}
(\nu^z)^{-1}\mathbf{1}=W(u^z_+,u^z_+)=W(u_-^zM^z_+\;+\;u_-^{\overz}N^z_+,u_-^zM^z_+\;+\;u_-^{\overz}N^z_+). 
\end{align}
Expanding the r.h.s. of \eqref{to2} and using Equations \eqref{w-asin-1}, \eqref{w-asin-2}, one gets
$$(\nu^z)^{-1}\mathbf{1}=(\nu^z) ^{-1}(M^z_+)^*M^z_+-(\nu^z)^{-1}(N^z_+)^*N^z_+,
$$ 
where $\nu^{1/z}=-\nu^z$ was used. This implies \eqref{I6}. Eq. \eqref{I2} is obtained in similar manner by expanding $W(u_-^z,u_-^z)$. Now let us prove $\eqref{I5}$.  It follows from Equations \eqref{w-asin-1} and \eqref{lineal} that
	\begin{align}\label{to3}
	0=W(u^{1/z}_+,u^z_+)=W(u_-^{1/z}M^{1/z}_+\;+\;u_-^{z}N^{1/z}_+,u_-^zM^z_+\;+\;u_-^{\overz}N^z_+).
	\end{align}
	Expanding the r.h.s. of \eqref{to3} and using Equations \eqref{w-asin-1}, \eqref{w-asin-2}, we get
	$$0=-(\nu^z)^{-1}(M_+^{1/z})^*N_+^z+(\nu^z)^{-1}(N_+^{1/z})^*M_+^z=-(\nu^z)^{-1}M_-^zN_+^z-(\nu^z)^{-1}N_-^{1/z}M_+^z,$$ where the last equality follows from \eqref{masamenos}. Eq. \eqref{I3} is obtained in similar manner expanding $W(u_-^{1/z},u_-^z)$. 
 \hfill $\Box$ 

\vspace{.2cm}

The next proposition allows to extend $M^z_{\pm}$ to $0$.  

\begin{proposi}\label{Men0}
The functions $z \mapsto M_{\pm}^z$ and $z \mapsto M_{\pm}^{1/z} $ are differentiable  on $\overline{\DM} \setminus \{-1,0,1 \} $ and $  \mathbb{S}^1 \setminus \{-1,1 \} $, respectively. Moreover, $$\lim_{z\to 0} M_{\pm}^{z} = \mathbf{1}.$$
 Therefore  the functions $z \mapsto M_{\pm}^z $ are analytic   on $|z|< 1$.
 \end{proposi} 

 \noindent {\bf Proof.} 
By Remark~\ref{deru} the functions $z\mapsto (u_-^{1/\overline{z}}(n))^*$, $z\mapsto u^z_+(n)$   are differentiable on $\DM\setminus\{-1,1\}$. By \eqref{M_+} and \eqref{M_-} this implies the first claim. The second part follows from the following computation using Eq. \eqref{M_-} 
  \begin{align*}
 M_-^z&=-\nu^zW(u_+^{\overline{z}},u_-^{1/z})
 \\
 & =\frac{z}{z^2-1}(u_+^{\overline{z}}(n+1)^*u_-^{1/z}(n)-u_+^{\overline{z}}(n)^*u_-^{1/z}(n+1))\\&=\frac{z^2}{z^2-1}(\overline{z}^{-(n+1)}u_+^{\overline{z}}(n+1))^*z^{n}u_-^{1/z}(n)-\frac{1}{z^2-1}(\overline{z}^{-n}u_+^{\overline{z}}(n))^*z^{n+1}u_-^{1/z}(n+1)\\&=\frac{z^2}{z^2-1}(\tilde{u}_+^{\overline{z}}(n+1))^*\tilde{u}_-^{1/z}(n)-\frac{1}{z^2-1}(\tilde{u}_+^{\overline{z}}(n))^*\tilde{u}_-^{1/z}(n+1) \to \mathbf{1} \;\;\mbox{ as } \ z \to 0,
 \end{align*}  
because $\tilde{u}_+^{z}(n),\tilde{u}_-^{1/z}(n) \to \mathbf{1}$ as $ z\to 0 $ (see Proposition \ref{jost-sol-1}). The other limit can be computed in the same fashion. The last claim follows from the removable singularity theorem.
 \hfill $\Box$

\vspace{.2cm}

Next recall Definition~\ref{disper} introducing the scattering matrix $\Ss^z$ for  $z\in\overline{\DM}$ by \eqref{eq-ScatMat}, provided that $M_{\pm}^z$ are invertible. This is the case on the unit circle:

\begin{proposi}\label{invertible}
	For $z\in \mathbb{S}^1\setminus\{1,-1\}$, the matrices  $M_{\pm}^z$ are invertible and the scattering matrix $\mathcal{S}^z$ is unitary. 
\end{proposi}
 \noindent {\bf Proof.} 
Eqs. \eqref{I2} and \eqref{I6} imply that  $M_{\pm}^z$ are invertible (using that $ \langle A^*A \phi, \phi \rangle = \| A \phi \|^2  $ one checks injectivity and therefore surjectivity because they are finite dimensional operators). For the second part, the off-diagonal terms of $  (\Ss^z)^*\,\Ss^z  $ are  (see Definition \ref{disper})
		\begin{align} 
		-((M_+^z)^{-1})^*N_{-}^z (M_-^z)^{-1} 
		-((M_+^z)^{-1})^*(N_{+}^{z})^* (M_-^z)^{-1} , \\
		-((M_-^{z})^{-1})^*(N_{-}^{z})^* (M_+^z)^{-1} - 
		((M_-^{z})^{-1})^* N_{+}^{z} (M_+^z)^{-1}
		\end{align}
and they vanish by \eqref{masamenos}. The  diagonal terms are 
		\begin{align}
		((M_+^z)^{-1})^* (1 + (N_+^z)^* N_+^z )  (M_+^z)^{-1} ,  \\ \notag
		(  (M_-^z)^{-1})^* (1 + (N_-^z)^* N_-^z )  (M_-^z)^{-1} ,
		\end{align}
and they are  both equal to  $\mathbf{1}$, see \eqref{I6} and \eqref{I2}.
This proves the unitary of $\Ss^z$.
 \hfill $\Box$

\section{Half-bound states}\label{halfbound-states}

This section analyzes the behavior of the function $z\mapsto \det(M^z_{+})$ when $z\to \pm1$. All the results in this section are presented for $z\to 1$, but they are also true for $z \to -1$ and the corresponding proofs are basically the same. Throughout this section we will denote $J_h^{+}=\dim\Ker(W(u_-^1,u_+^1))$. Notice that this seems to differ from Theorem \ref{theo-Levinson}, however, these two definitions coincide as will be verified in the proof of Theorem \ref{theo-Levinson}. Let us start by  stating a result from \cite{BFGS} (see Proposition 24 and Eq. (102) therein). From the  definition of $u_+^1$, one knows that $u_{+}^1(j)$ tends to {${\bf 1}$}  
as $j$ tends to infinity. Then, for large enough $j$,    $u_{+}^1(j)$ is invertible.  In order to simplify notations, we assume that   $u_{+}^1(1)$  is already invertible (this is needed in order to apply Proposition 24 in \cite{BFGS}). This does not imply any restriction because one can always  translate the origin.

\begin{proposi}\label{asin_W}
	There exist invertible matrices $P, Q\in \mathcal{M}_{L\times L}$ and matrix valued functions $A(z), B(z), C(z), D(z)$,  for $z \in \overline \DM\setminus \{ 1 \} $ (recall \eqref{disc}) sufficiently close to $1$,  such that 
	\begin{equation}\label{piri}
	P \ W(u_{-}^{1/\overline{z}}, u_{+}^z)
	  u_+^z(1)^{-1}u_+^1(1)Q  = \begin{pmatrix}
	A(z)  & B(z) \\ C(z)  & D(z)
	\end{pmatrix}
	\end{equation}
	where
\begin{equation}
\label{mainnnn}
A(z) = \imath (1 - z) \boldsymbol{ A} + o(|1-z|),  \quad B(z) = o(1), \quad C(z) = \Oo(|1-z|),  \quad D(z) = \boldsymbol{ D} + o(1).
\end{equation}
In the previous equations, $\boldsymbol{ A}$ is a matrix of size $J_h^+\times J_h^+$ (and this determines the dimensions of the other matrices involved). Moreover,  $\boldsymbol{ A}$ and 
and $ \boldsymbol{ D}$ are invertible matrices. The invertibility of $  u_+^z(1) $ for $z$ close to $1$ follows from the invertibility of $u_+^1(1)$ and the continuity of Jost solutions.   
\end{proposi} 

In \cite{BFGS}, Proposition \ref{asin_W} was used to show that the limits ${T_{\pm}^1:=}\lim_{z \to 1}  T_{\pm}^z $ exist. It also implies the next result which is a preparation for the proof of Levinson's theorem. 

\begin{proposi}\label{anal_1}
	There is a constant $c \in \mathbb{C}\setminus\{0\} $ such that 
$$\det(M_+^z)=(z-1)^{J_h^{+}-L}(c+o(1)) , \qquad  z \to  1 \;,\;\; z \in \overline{\mathbb{D}}. 
$$
\end{proposi}
	\noindent {\bf Proof.} Using Eqs. \eqref{M_+} and \eqref{piri} one has for $0<|z|\leq1$ that
	\begin{equation}\label{nose}
		\det(M_+^z)=(\nu^z)^L\det \begin{pmatrix}
		A(z)  & B(z) \\ C(z)  & D(z)
		\end{pmatrix}(a+o(1)),
	\end{equation} where $a=\det(PQ)^{-1} \neq 0$, and the continuity of the function $z\mapsto u^z_+(1)$ was used. Using Schur formula  for the determinant (see Proposition~\ref{schur}) and Eq. \eqref{mainnnn}, it follows that
	\begin{equation}\label{nose_2}
		\det \begin{pmatrix}
		A(z)  & B(z) \\ C(z)  & D(z)
		\end{pmatrix} = \det(\boldsymbol{ D} + o(1)) \det(\imath (1 - z) \boldsymbol{ A} + o(|1-z|))=(z-1)^{J_h^+}(b+o(1)),
	\end{equation} 
where $b$ is a non-zero constant. Using Eqs. \eqref{nose}, \eqref{nose_2} and \eqref{lanu}, the required result follows.
\hfill $\Box$ 

\begin{rem}\label{para_-}
{\rm Using Eq. \eqref{masamenos} and Proposition \ref{anal_1} one gets a similar result for $M_-^z$ in  a neigh\-bor\-hood of $z = 1$ in $ \overline{\mathbb{D}} $:
	\begin{equation}\label{halfM+}
		\det(M_-^z)=(z-1)^{J_h^+-L}(\overline{c}+o(1)).
	\end{equation}
The corresponding result in a neighborhood of $z=-1$ in $ \overline{\mathbb{D}} $ reads as:
	\begin{equation}\label{halfM-}
		\det(M_-^z)=(z+1)^{J_h^--L}(d+o(1)),
	\end{equation}
where $d\in \mathbb{C}\setminus\{0\}$. 
}
\hfill $\diamond$
\end{rem}

 \section{Bound states}\label{bound-states}

This section is about the behavior of the function $z\mapsto \det(M_{\pm}^z)$ when $z \to r$, where  $r$ is such that $E=r+1/r$ is an eigenvalue of $H$. The main result is (see Proposition \ref{multiplicity-bounded}) that the number of zeros of the function $z \mapsto \det(M^z_{\pm})$ on $\mathbb{D}$ (counted with multiplicity)   equals the number of eigenvalues of $H$ (counted with multiplicity).  
 
\begin{proposi}\label{boundstates}
   For $z\in \mathbb{C}$, $0<|z|<1$, and $E=z+1/z$, the following identity holds true:
   \begin{equation}\label{dimen}
\dim(\Ker(H-E))=\dim(\Ker( M_{\pm}^z)). 
   \end{equation}
Moreover,  $N_-^z$ restricted to $\Ker( M_-^z)$ is a bijection between $\Ker (M_-^z)$ and $\Ker( M_+^z)$.
 \end{proposi}

 \noindent {\bf Proof.} Let us prove \eqref{dimen} for $M_+^z$ and the result for $M_-^z$ is obtained in a similar fashion. Set 
 $$\mathcal{S}_+=\{\phi \in \mathbb{C}^L: u_+^z\phi\in \ell^2(\mathbb{Z}, \mathbb{C}^L)\}, \qquad 
 \mathcal{S}_-=\{\phi \in \mathbb{C}^L: u_-^{1/z}\phi\in \ell^2(\mathbb{Z}, \mathbb{C}^L)\},
 $$ 
 then the function $T:\mathcal{S}_+ \to \Ker(H-E)$ defined by $T(\phi)=u_+^{z}\phi$ is linear and injective (because the columns of $u_+^{z} $ are linearly independent by Remark \ref{fun_sol} and these columns are precisely solutions to the eigenvalue problem). Let $u \in \Ker(H-E)$, then there exist $\phi \in \mathbb{C}^L$ such that
 $ u(n)=u_+^{z}(n)\phi$ (write $u=u_+^{z}\phi +u_+^{1/z}\psi$ again by Remark~\ref{fun_sol} and notice that  $u_+^{1/z}(n) \psi \ne 0  $ implies that $ \lim_{n \to +\infty }\|u_+^{1/z}(n) \psi  \|  = \infty $  - see Eq. \eqref{u-asin}). Thus  $T$ is surjective, and it is consequently an isomorphism.

\vspace{.1cm}
 
Next  we prove that $\mathcal{S}_+ = \Ker(M_+^z)$ which implies   \eqref{dimen} (similarly,  one proves  that $\mathcal{S}_- = \Ker(M_-^z)$). Let us take $\phi \in \Ker(M_+^z)$ and multiply  \eqref{lineal} by $\phi$ so that
 \begin{equation}\label{tra}
   u_+^{z}\phi= u_-^{1/z}N_+^z\phi. 
 \end{equation}
This implies that   $u_+^{z}\phi \in \ell^2(\mathbb{Z},\mathbb{C}^L)$ and therefore  $\phi \in \mathcal{S}_+$, which implies that
 $ \Ker(M_+^z) \subset \mathcal{S}_+  $.   The other contention is proved by taking $\phi \in \mathcal{S}_+$ and multiplying \eqref{lineal} by $\phi$. Then
$$u_+^z\phi= u_-^{z}M_+^z\phi + u_-^{1/z}N_+^z\phi. 
$$
Since $u_+^{z}\phi\in \ell^2(\mathbb{Z},\mathbb{C}^L)$, it follows (using the asymptotic behavior of Jost solutions to compute the second term on the right of the next equation)  that
 $$
 \lim_{n\to -\infty}u_{-}^{z}(n)M_+^z\phi = \lim_{n\to -\infty} u_+^{z}(n)\phi - u_-^{1/z}(n)N_+^z\phi=0.
 $$ 
The asymptotic behavior of Jost solutions implies that $M_+^z\phi=0$ (since otherwise  one would have $\lim_{n \to -\infty}\|u_{-}^{z}(n)M_+^z \phi \| = \infty $). The arguments above imply the fist part of the statement. 

\vspace{.1cm}
 
Next, let us prove that $N_-^z\big|_{\Ker( M_-^z)}$ is a bijection between $\Ker( M_-^z)$ and $\Ker (M_+^z)$. 
 Take $\phi \in\Ker(M_-^z)$.  Eq. \eqref{lineal} implies that 
$$u_-^{1/z}\phi = u_+^z N_-^z\phi,
$$
and using the asymptotic behavior of Jost solutions  (see Eq. \eqref{u-asin}) one  concludes that   
$$
u_+^z N_-^z\phi \in \ell^2(\mathbb{Z},\mathbb{C}^L).
$$  
With help of Eq. \eqref{lineal} for $ u_+^z  $ ({\it i.e.}  $ u_+^z N_-^z\phi = u_-^z M_+^z N_-^z\phi + u_{-}^{1/z} N_+^z N_-^z\phi  $ ), one deduces as before (using a blow up argument) that  $N_-^z\phi \in \Ker(M_+^z)$. This implies that $ N_-^z $ maps $\Ker (M_-^z)$ into $ \Ker( M_+^z )$. Moreover,  Eq. \eqref{lineal} and the above equations imply that 
$$ 
u_-^{1/z}\phi = u_+^zN_-^z\phi=u_-^{1/z}N_+^zN_-^z\phi
.
$$
Taking the limit $n \to -\infty$ in this identity  (see also Eq. \eqref{u-asin}), it follows that
$$
\phi =N_+^zN_-^z\phi.
$$ 
In a similar fashion, one proves that if $\phi \in \Ker(M_+^z)$ then $N_+^z\phi \in \Ker(M_-^z)$ and 
$$\phi =N_-^zN_+^z\phi.
$$ 
Then  the restriction of $ N_+^z $ to $\Ker(M_+^z)$  is the inverse of $N_-^z\big|_{\Ker (M_-^z)}$, concluding the proof.\hfill $\Box$

\begin{proposi}\label{finite-spectrum}
  The set of eigenvalues of $H$ is finite and every eigenvalue $E$ can be expressed in the form 
  $$
  E = z + 1/z,
  $$    
  for some  $  z \in  (-1, 0) \cap (0,1) $. 
\end{proposi}
 
  \noindent {\bf Proof.} 
Let us first state some properties of the function
$$
E: \mathbb{C} \setminus \{0 \}  \to 
\mathbb{C},  \qquad  E(z) := z + 1/z: 
$$ 
Solving for $z$ gives
$$
z = E/2 + \sqrt{  E^2/4   -1  },
$$
which implies that the map $E$ is surjective. The presence of the square root  implies that the solutions are given by a Riemann surface with two branches. Then,  for every complex number $ \boldsymbol{E}$, there are only two solutions $z_{in},  z_{ext}$ for the equation 
 $$
 E(z) =  \boldsymbol{E}. 
  $$  
Since  $ E(z) = E(1/z) $, one obtains that the restriction of $E$ to the disc  $ \DM $ is injective and its restriction to $ \overline \DM $ is surjective (and therefore the analysis can be restricted to the case $|z|\leq 1$).  
Moreover,  an elementary calculation yields that the equation 
$$z+1/z=E,$$ 
for  $E$ in the real numbers and $|z|\leq 1$, is solvable  if and only if $|z| = 1$ or $  z\in (-1,1)  $.
 
\vspace{.1cm}

Since $H$ a self-adjoint, its spectrum is contained in the real line. All eigenvalues, parameterized in the form $E(z)$,  $|z| \leq 1$, must satisfy that $z \in \mathbb{S}^1 \cup (-1,1) $.  Next let us argue that, furthermore, if $z \in \mathbb{S}^1 \setminus \{  1, -1 \}$, then $E = z+ 1/z$ is not an eigenvalue. Suppose that $u$ is an eigenvector of $H$ corresponding to $E(z)$, {\it i.e.} $H u = E(z)u$, with $  z \in \mathbb{S}^1 \setminus \{-1, 1 \} $.  Remark~\ref{fun_sol}  implies that $u$ can be written in the  form 
$$u= u_+^z   \alpha+ u_{+}^{1/z}\beta , $$ 
for some $\alpha,\beta \in \mathbb{C}^L$.
As $ u$ is square integrable, one has 
$$ \lim_{n \to \infty} u(n)  = 0.$$  
Eq.  \eqref{u-asin} yields that 
$$ \lim_{n \to 
\infty     }   z^n \alpha+ (1/z)^n \beta  = 0,$$ 
which is only possible when $ \alpha = \beta = 0$ and hence $u=0$.
Consequently  all eigenvalues must lie  on $[-1, 1]$. It remains to rule out the points $\{-1, 1  \}$. We analyze 
   only   $ z = 1 $, since
    the analysis for $z = -1 $ 
   is the same.  The proof in this 
   case 
   is similar, but   Remark 
   \ref{fun_sol} is not valid 
   anymore because $u_{+}^{z} = u_{+}^{1/z} $ for
    $z= 1$ . The columns of $  u_{+}^{ 1} $ 
    do not generate all solutions. Nevertheless, in Definition 1 in  \cite{BFGS} we introduce another solution $ v_{+}^{1}    $ such that the columns of $ [   u_{+}^{1}  \:   v_{+}^{1}   ]$ generate all solutions. Now, following the 
         line-of-argument  for the case $ z \in \mathbb{S}^1\setminus \{ -1, 1 \} $, one concludes that  $1$ is not an eigenvalue. 

\vspace{.1cm}

Next let us check that there are neighborhoods of $  0 $ and $ \pm 1 $ in $ \overline{\DM}  $  such that for $z$ in these neighborhoods $ E(z) $  is not an eigenvalue of $H$. Proposition~\ref{boundstates} implies that a number  $  E=z+1/z\in \mathbb{C}$ with $0<|z|<1$ is an eigenvalue of $H$  if and only if the function $z' \mapsto \det(M_{\pm}^{z'})$ has a zero at $z$. Now $M^{z'}_\pm$ is invertible in a neighborhood of $\pm 1$  (by Proposition \ref{anal_1} and Remark \ref{para_-}) and also in a neighborhood of $0$ by Proposition~\ref{Men0}. This implies the claim.

\vspace{.1cm}

Finally let us recall that the essential spectrum of 
$ H  $ is $[ -2, 2]$ which is precisely the image of $\SM^1$ under the map $E$.  The above arguments imply that the eigenvalues of $ H $  take the form $ E(z) $, for $z$ in a compact subset $ K $ of $  (0, 1) \setminus \{ 0 \}  $. Then, all eigenvalues of $H$ must belong to the compact set $E (K)$.  This set does not intersect the essential spectrum of $H$. Since all spectral points of $H $ not belonging to the essential spectrum are isolated eigenvalues with finite multiplicity, we conclude that there is only a finite  number of them (and they can be parametrized in the form $E(z)$ for  a finite number of $z$'s  in $(-1,0 ) \cup (0,1)$).          
 \hfill $\Box$

\vspace{.2cm}

Proposition \ref{boundstates} claims that the number of zeros, counted without multiplicity, of the function $z \mapsto \det(M_{\pm}^z)$ on $\DM$ is equal to the number of eigenvalues of $H$, counted without multiplicity. Proposition \ref{multiplicity-bounded} below  proves that they are also the same if counted with multiplicity. For the proof, the following technical statement is needed which is a discrete version of a result from \cite{AW0} that was already used in \cite{BFS}.

\begin{lemma}\label{tuncay}
  Let $r\in \mathbb{R}$, with $0<|r|<1$ and such that $r+1/r=E$ is an eigenvalue of $H$. Let $\alpha \in \Ker(M_-^r)$. The following equation holds true:  
  \begin{equation}\label{nor}
     (N_-^r\alpha)^*\frac{d}{dz}M_-^z\Big|_{z=r}\alpha=r^{-1}\|u^{1/r}_-\alpha\|^2
  \end{equation}
\end{lemma}
 \noindent {\bf Proof.} 
Let $z\in \DM$ and recall that   the Jost solution $ u^z_+ $ satisfies the generalized eigenvalue equations $H u^z_+ = E u^z_+ $ with $E = z+1/z$, namely
  \begin{align}\label{noma} u^z_+(n+1)+u^z_+(n-1)+V(n)u^z_+(n)=(z+1/z)u^z_+(n),   \qquad  \forall n \in \mathbb{Z}.
  \end{align}
  Taking derivative w.r.t. $r$, one obtains
  \begin{equation}\label{der}
    \dot{u}^r_+(n+1)+\dot{u}^r_+(n-1)+V(n)\dot{u}^r_+(n)=(r+1/r)\dot{u}^r_+(n)+(1-1/r^2)u^r_+(n),
  \end{equation}
  where  $\dot{u}^r_+ $ is given by $ \dot{u}^r_+(n)= \frac{d}{dz}u_+^z(n)\Big|_{z=r}$. Taking adjoints and evaluating in $z=r$ in \eqref{noma} leads to
  \begin{equation}\label{adj}
    u^r_+(n+1)^*+u^r_+(n-1)^*+u^r_+(n)^*V(n)=(r+1/r)u^r_+(n)^*.
  \end{equation}
  Multiplying \eqref{der} on the left by $u^r_+(n)^*$ and \eqref{adj} on the right by  $\dot{u}^r_+(n)$ and subtracting the  resulting equations, one obtains
  \begin{align} \label{noma1} u^r_+(n)^*\dot{u}^r_+(n+1)+u^r_+(n)^*\dot{u}^r_+(n-1)-u^r_+(n+1)^*\dot{u}^r_+(n)- & u^r_+(n-1)^*\dot{u}^r_+(n) \\ \notag  & =(1-1/r^2)u^r_+(n)^*u^r_+(n).
  \end{align}
Recalling  the  definition of the Wronskian $W(u^r_+,\dot{u}^r_+)(n)=\imath (u^r_+(n+1)^*\dot{u}^r_+(n)-u^r_+(n)^*\dot{u}^r_+(n+1))$,  one can  rewrite the last equation as
  $$W(u^r_+,\dot{u}^r_+)(n-1)-W(u^r_+,\dot{u}^r_+)(n)=\imath(1-1/r^2)u^r_+(n)^*u^r_+(n).$$
  Multiplying this equation by $N_-^r\alpha$ on the right and by  $(N_-^r\alpha)^*$ on the left implies that
  \begin{equation}\label{mul}
    (N_-^r\alpha)^*(W(u^r_+,\dot{u}^r_+)(n-1)-W(u^r_+,\dot{u}^r_+)(n))N_-^r\alpha=i(1-1/r^2)(u^{1/r}_-(n)\alpha)^*u^{1/r}_-(n)\alpha,
  \end{equation}
where Eq. \eqref{lineal} was used to exchange $u^r_+(n)N_-^r\alpha$ by $u^{1/r}_-(n)\alpha$ (recall that $\alpha \in \Ker(M_-^r)$). Since $\alpha \in \Ker(M_-^z)$, one has that $u^{1/r}_-\alpha\in \ell^2(\mathbb{Z},\mathbb{C})$ (by using that $u^r_+(n)N_-^r\alpha = u^{1/r}_-(n)\alpha$ and the asymptotic properties of Jost solutions). Now  take the sum in both sides of the Eq. \eqref{mul} to get:
  $$\sum_{n\in \mathbb{Z}}  s(n-1)-s(n) =  \sum_{n \in \mathbb{Z}}i(1-1/r^2)(u^{1/r}_-(n)\alpha)^*u^{1/r}_-(n)\alpha=i(1-1/r^2)\|u^{1/r}_-\alpha\|^2,$$
  where $s(n):=(N_-^r\alpha)^*W(u^r_+,\dot{u}^r_+)(n)N_-^r\alpha, \ n \in \mathbb{Z}$.
  Note that the l.h.s.  of the equation is a telescoping series. Thus
  \begin{equation}\label{limit}
  	\lim_{n\to -\infty}s(n)-\lim_{n\to +\infty}s(n)=i(1-1/r^2)\|u^{1/r}_-\alpha\|^2. 
  \end{equation}
Calculating $\dot{u}^r_+(n)= nr^{n-1}\tilde{u}_+^r(n)+r^n \frac{d}{dz} \tilde u_+^z(n)$ and noticing that $\frac{d}{dz} \tilde u_+^z,\tilde u_+^z \in \ell^\infty(\mathbb{N},\Mm)$  (see Proposition \ref{derdtu}), one obtains that   
  \begin{align}\label{noma2}     u^r_+(n) , \qquad  \dot{u}^r_+(n) \to 0, n\to +\infty.
  \end{align} 
 Thus $$ s(n) \to 0\;\;\mbox{ as } n \to +\infty  .$$ Using the definition of the Wronskian (see Eq. \eqref{wronskian}) and the general fact that $\frac{d}{dz}f(\overline{z})^*\Big|_{z=z_0}={(\frac{d}{dz}f(z)\Big|_{z=\overline{z_0}})}^*$, one obtains the following (for every $n \in \mathbb{Z}$): 
   \begin{equation}\label{deriwron}
  	\frac{d}{dz}W(u_+^{\overline{z}},u_-^{1/z})\Big|_{z=r} = W(\dot{u}_+^r,u_-^{1/r})(n)+  W(u_+^r,\dot{u}_-^{1/r})(n),
  \end{equation}  
because $\dot{u}_-^{1/r}$ is given by $ \dot{u}_-^{1/r}(n)= \frac{d}{dz}u_-^{1/z}(n) \Big|_{z=r}.$ 
The following computation now uses again Eq. \eqref{lineal} in order  to replace $u^r_+(n)N_-^r\alpha$ by $u^{1/r}_-(n)\alpha$ (recall that $\alpha \in \Ker(M_-^r)$) and Eq. \eqref{deriwron} 
  \begin{equation*}
    \begin{aligned}
       s(n)&=(N_-^r\alpha)^*W(u^r_+,\dot{u}^r_+)(n)N_-^r\alpha=W(u^r_+N_-^r\alpha,\dot{u}^r_+)(n)N_-^r\alpha=W(u^{1/r}_-\alpha,\dot{u}^r_+)(n)N_-^r\alpha\\ \\ &= \alpha^*W(u^{1/r}_-,\dot{u}^r_+)(n)N_-^r\alpha
       = \alpha^*\left(	\frac{d}{dz}W(u_+^{\overline{z}},u_-^{1/z})\Big|_{z=r} - W(u_+^r,\dot{u}_-^{1/r})(n)\right)^*N_-^r\alpha= \\ \\
       &=\alpha^*(\frac{d}{dz}W(u_+^{\overline{z}},u_-^{1/z})\Big|_{z=r})^*N_-^r\alpha - \alpha^*W(\dot{u}_-^{1/r},u_-^{1/r})(n) \alpha,
    \end{aligned}
  \end{equation*}
  Arguing as in
   \eqref{noma2}, one gets $\dot{u}_-^{1/r}(n),u_-^{1/r}(n)\to 0, \ n \to -\infty.$  Taking the limit $(n \to - \infty)$ in the last equation leads to
  \begin{equation}\label{chateu}
  	s(n)^* \to (N_-^r\alpha)^*\frac{d}{dz}W(u_+^{\overline{z}},u_-^{1/z})\Big|_{z=r}\alpha , \ n \to -\infty. 
  \end{equation}
Eqs. \eqref{limit} and \eqref{chateu} and the fact that  $s(n) \to 0, \ n \to +\infty,$ show that 
   \begin{equation}\label{igualdad_1}
		 (N_-^r\alpha)^*\frac{d}{dz}W(u_+^{\overline{z}},u_-^{1/z})\Big|_{z=r}\alpha = -i(1-1/r^2)\|u^{1/r}_-\alpha\|^2.
   \end{equation} Using Eq. \eqref{M_-} implies
  \begin{equation}\label{der_M_-}
  	\frac{d}{dz}M_-^z\Big|_{z=r}=-\frac{d}{dz}\nu^z\Big|_{z=r}W(u_+^r,u_-^{1/r})-\nu^z\frac{d}{dz}W(u_+^{\overline{z}},u_-^{1/z})\Big|_{z=r}.
  \end{equation} Then,  due to $\alpha \in \Ker(M_-^z)= \Ker W(u_+^r,u_-^{1/r})$,
  \begin{equation}\label{der_Mult}
  (N_-^r\alpha)^*\frac{d}{dz}M_-^z\Big|_{z=r}\alpha=-\nu^z(N_-^r\alpha)^*\frac{d}{dz}W(u_+^{\overline{z}},u_-^{1/z})\Big|_{z=r}\alpha.
  \end{equation} 
Combining Eqs. \eqref{der_Mult} and \eqref{igualdad_1}, the result follows.
 \hfill $\Box$ 
   
\begin{proposi}\label{multiplicity-bounded}
  Suppose that $r\in \mathbb{R}$ with $0<|r|<1$ and 
  that $E=r+1/r$ is an eigenvalue of $H$.  Set $n_r=\dim(\Ker( M_-^r))=\dim (\Ker (H-E))$.  Then there exists a complex number   $c_r\in \mathbb{C}\setminus\{0\}$ such that 
  $$\det(M_{-}^{z})=(z-r)^{n_r}(c_r+  \Oo(|z-r|) ), \qquad  z \to r. $$
\end{proposi}

 \noindent {\bf Proof.}  Let $\{u_1,...,u_{n_r}\} $ be a  basis of $\Ker(M_-^r)$. Since $N_-^r\big|_{\Ker(M_-^r)}:\Ker( M_-^r) \to \Ker (M_+^r) $ is an isomorphism (see Proposition \eqref{boundstates}),   it follows that $\{N_-^ru_1,...,N_-^{r}u_{n_r}\}$ is a basis of $\Ker(M_+^r)=\Ran(M_-^r)^\perp$, the latter due to Eq. \eqref{masamenos}. Next let $\{v_{n_{r}+1},...,v_{L}\}$ be an orthonormal  basis of $\Ran(M_-^r)$ and $\{u_{n_{r}+1},...,u_{L}\}$ such that $M_-^ru_i=v_i$. Then $\{N_-^ru_1,...,N_-^{r}u_{n_r},v_{n_{r+1}},...,v_{L}\}$ and $\{u_1,...,u_{n_r},u_{n_{r}+1},...,u_{L}\}$  are basis of $C^L$. We denote by $U_1,V_1 $ and $V_1,V_2 $ the matrices such that $$U_1=(u_1...u_L), \ U_2=(u_1...u_{n_r}), \ \ V_1=(N_-^ru_1 ...N_-^{r}u_{n_r} \ v_{n_{r+1}}...v_{L}), \  V_2=(N_-^ru_1 ...N_-^{r}u_{n_r}).$$ Then  
    $$V_1^*M_-^rU_1=\begin{pmatrix}
    0 & 0\\
    0 & 1
    \end{pmatrix}.$$ 
    We let $\tilde A, \tilde B,\tilde  C, \tilde D$ denote the matrices satisfying  
 $$ 
    \ \ \ V_1^*\frac{d}{dz}M_-^z\Big|_{z=r}U_1=\begin{pmatrix}
    	\tilde A & \tilde  B \\
    \tilde 	C & \tilde D
    \end{pmatrix},
 $$   
  where  $\tilde A=V_2^*\frac{d}{dz}M_-^z\Big|_{z=r}U_2$.   Lemma \ref{tuncay} shows that $\tilde A$ is invertible because for $\phi \in \mathbb{C}^{n_r}\setminus\{0\}$ 
    $$\phi^* \tilde A\phi = (V_2\phi)^*\frac{d}{dz}M_-^z\Big|_{z=r}U_2\phi =(N_-^rU_2\phi)^*\frac{d}{dz}M_-^z\Big|_{z=r}U_2\phi=r^{-1}\|u_-^{1/r}U_2\phi\|^2\neq 0.$$
Note that the last identity used that the columns of $U_2$ are linearly independent. This implies that  $U_2 \phi \neq 0 $. The fact that $ \|u_-^{1/r}U_2\phi\|^2\neq 0 $  follows from  Eq. \eqref{u-asin}, which implies that if $x\neq 0$ then  $r^n u_-^{1/r}(n)x\to x \neq 0$ as $ n \to -\infty$.   With the help of  Taylor's theorem and analyticity, it follows that  
    \begin{equation}\label{M_-^z}
    \begin{aligned}
    		V_1^*M_-^zU_1&=\begin{pmatrix}
    	0 & 0\\
    	0 & 1
    	\end{pmatrix}+(z-r)V_1^*\frac{d}{dz}M_-^z\Big|_{z=r}U_1+\Oo((z-r)^2)\\&=\begin{pmatrix}
    	(z-r) \tilde A & (z-r) \tilde B\\
    	(z-r) \tilde C & 1+(z-r) \tilde D
    	\end{pmatrix}+\Oo((z-r)^2)\;\; \mbox{ as } z\to r.
    \end{aligned}    
    \end{equation}
    Using the Schur formula (see Proposition~\ref{schur}) for the determinant in Eq. \eqref{M_-^z} one gets 
      \begin{equation*}
    	\begin{aligned}
    		\det(V_1^*M_-^zU_1)&=\det(1+(z-r) \tilde D +  \Oo((z-r)^2 ))\cdot \\ & \hspace{2cm}\det((z-r)\tilde A+ (z-r)^2\tilde B(1+o(1))\tilde C+\Oo((z-r)^2) )\\&=\det(1+(z-r) \tilde D+\Oo((z-r)^2))\det((z-r)(\tilde A+\Oo(z-r)))\\&=(z-r)^{n_r}g(z),
    	\end{aligned}
    \end{equation*}
where $g(z)=\det(1+(z-r)D+\Oo(z-r)^2)\det(\tilde A+\Oo(z-r))$. From the last equation and the fact that $g(r)=\det(\tilde A)\neq0$  the desired result follows.
 \hfill $\Box$ 
     
\section{Time delay}\label{STime}

The (total) time delay is by definition the quantity
$$
\Tr\Big((\Ss^z)^*\frac{d}{dz} \Ss^z\Big)
\;=\;
\det(\mathcal{S}^z)^{-1}\frac{d}{dz}\det(\mathcal{S}^z)
$$
for $z\in\SM^1\setminus\{-1,1\}$ (the above identity is referred to as Jacobi's formula). This section provides a formula for it in terms of the determinant of $M^z_-$.

\begin{proposi}\label{split}
	Let $z\in \mathbb{S}^1\setminus\{1,-1\}$.  The following identity holds true:
	$$\det(\mathcal{S}^z)=\det(M_-^z)^{-1}\det((M_+^z)^*)=\det(M_-^z)^{-1}\det(M_-^{1/z}),$$
\end{proposi}
 \noindent {\bf Proof.} 
Applying the Schur complement formula for the determinant (see Proposition~\ref{schur}) to the definition \eqref{eq-ScatMat} of the scattering matrix leads to
	\begin{equation}\label{tacos_canasta}
		\det(\mathcal{S}^z)=\det(M_-^z)^{-1}\det((M_+^z)^{-1}-N_-^zN_+^z(M_+^z)^{-1}).
	\end{equation}
	Using Eqs. \eqref{masamenos} and  \eqref{I6} one obtains that
	\begin{equation}\label{salsa_verde}
	\begin{aligned}
		\det((M_+^z)^{-1}-N_-^zN_+^z(M_+^z)^{-1})&=\det((\mathbf{1}-N_-^zN_+^z)(M_+^z)^{-1})
		=\det((\mathbf{1}+(N_+^z)^*N_+^z)(M_+^z)^{-1})\\&=\det((M_+^z)^*)=\det(M_-^{1/z}).		
	\end{aligned}		
	\end{equation}
	Equations \eqref{tacos_canasta} and \eqref{salsa_verde} imply the claim.
 \hfill $\Box$ 

\vspace{.2cm}

Propositions \ref{split} and \ref{Men0} imply that 
the function $z  \mapsto \det(\mathcal{S}^z) $ is differentiable on $ \mathbb{S}^{1}\setminus \{-1,1 \}.  $  This allows us to state the next result.

\begin{coro}\label{time_delay}
	For every $z\in \mathbb{S}^1\setminus\{1,-1\}$, the following hold true: 
	\begin{equation}\label{time_delay_eq}
			\det(\mathcal{S}^z)^{-1}\frac{d}{dz}\det(\mathcal{S}^z)=\det(M_-^{1/z})^{-1}\frac{d}{dz}\det(M_-^{1/z}) - \det(M_-^z)^{-1}\frac{d}{dz}\det(M_-^z). 
	\end{equation}
\end{coro}
 \noindent {\bf Proof.} 
	 Using Proposition \ref{split}, an explicit computation gives 
	\begin{equation}\label{deri_matrix}
		\frac{d}{dz}\det(\mathcal{S}^z)=\det(M_-^z)^{-1}\frac{d}{dz}\det(M_-^{1/z})-\det(M_-^z)^{-2}\det(M_-^{1/z})\frac{d}{dz}\det(M_-^z). 
 \end{equation}
	Multiplying Eq. \eqref{deri_matrix} by $\det(\mathcal{S}^z)^{-1}=\det(M_-^z)\det(M_-^{1/z})^{-1}$ one gets the stated result. 
 \hfill $\Box$ 

\section{Proof of Levinson's theorem}\label{Slevinson}

  \noindent {\bf Proof}  of Theorem~\ref{theo-Levinson}. For each $\epsilon>0$, let $\Gamma_+^\epsilon$ and $\Gamma_-^\epsilon$ be the truncated upper and lower semicircles parameterized by $\gamma_+^\epsilon,\gamma_-^\epsilon:[0,1] \to \mathbb{S}^1,$  $$\gamma_+^\epsilon(t)=e^{\imath\pi((1-t)\epsilon+t(1-\epsilon))}, \ \ \ \gamma_-^\epsilon(t)=1/\gamma_+^\epsilon(1-t). $$  
For every  $\delta>0$, let us denote by  $\Omega_+^{\epsilon,\delta} $ and  $\Omega_-^{\epsilon,\delta}$ the  interior arcs parameterized by $\omega_+^{\epsilon,\delta},\omega_-^{\epsilon,\delta}     [0,1] : \to \mathbb{C},$ given by 
$$\omega_+^{\epsilon,\delta}=(1-\delta)\gamma_+^\epsilon, \ \ \ \omega_-^{\epsilon,\delta}=(1-\delta)\gamma_-^\epsilon.$$
We let $l_+^{\epsilon,\delta}$ be the line segment  from $\omega_-^{\epsilon,\delta}(1)$  to $\omega_+^{\epsilon,\delta}(0)$,  and $l_-^{\epsilon,\delta}$  the line segment that goes from $\omega_+^{\epsilon,\delta}(1)$ to $\omega_-^{\epsilon,\delta}(0)$.  Now we define the positively-oriented  closed curve $\Omega_{\epsilon,\delta}:=\Omega_+^{\epsilon,\delta}+l_+^{\epsilon,\delta}+\Omega_-^{\epsilon,\delta}+l_-^{\epsilon,\delta}.$  By Propositions \ref{Men0}, \ref{boundstates}, \ref{finite-spectrum}, \ref{multiplicity-bounded} and  the argument principle one has that
 \begin{equation}\label{J_b}
 \lim\limits_{\epsilon \to 0} \lim\limits_{\delta \to 0} \int_{\Omega_{\epsilon,\delta}} \det(M_-^z)^{-1}\frac{d}{dz}\det(M_-^z) dz= 2\pi \imath J_b.
 \end{equation}
On the other hand,  Eq. \eqref{halfM+} implies that 
\begin{equation*}
\frac{d}{dz}\det(M_-^z)= (z-1)^{J_h^+-L}g'(z)+ (J_h^+-L)(z-1)^{J_h^+-L-1}g(z), 
\end{equation*}
where $g(z) \to \overline{c} \neq 0, \ z\to 1.$ Then  
\begin{equation}\label{derlog}
 \det(M_-^z)^{-1}\frac{d}{dz}\det(M_-^z)=(J_h^+-L)\frac{1}{z-1} + \frac{g'(z)}{g(z)}.
\end{equation} 
Using \eqref{derlog} and Lemma \ref{aux}, one can compute the next limit
\begin{equation}\label{J_h_+}
\begin{aligned}
\lim\limits_{\epsilon \to 0} \lim\limits_{\delta \to 0} \int_{l_+^{\epsilon,\delta}} \det(M_-^z)^{-1}\frac{d}{dz}\det(M_-^z) dz &=\lim\limits_{\epsilon \to 0} \lim\limits_{\delta \to 0} \int_{l_+^{\epsilon,\delta}}(J_h^+-L)\frac{1}{z-1} + \frac{g'(z)}{g(z)} dz\\&= (J_h^+-L) \lim\limits_{\epsilon \to 0} \lim\limits_{\delta \to 0} \int_{l_+^{\epsilon,\delta}}\frac{1}{z-1}dz =
{-
 \pi \imath (J_h^+-L). }
\end{aligned}	
\end{equation} An analogous calculation using \eqref{halfM-}, shows that 
\begin{equation}\label{J_h_-}
	\lim\limits_{\epsilon \to 0} \lim\limits_{\delta \to 0} \int_{l_-^{\epsilon,\delta}} \det(M_-^z)^{-1}\frac{d}{dz}\det(M_-^z) dz =
{-	
	 \pi \imath (J_h^--L)}.
\end{equation}
Now Lemma ~\ref{volteo} implies that 
\begin{equation}\label{int_p_p}
\begin{aligned}
		\int_{\Gamma_-^\epsilon}\det(M^z_-)^{-1}\frac{d}{dz}\det(M_-^z)dz=-\int_{\gamma_+^\epsilon} \det(M_-^{1/z})^{-1}\frac{d}{dz}\det(M_-^{1/z})dz.
\end{aligned}
\end{equation} 
Using the previous equation and \eqref{time_delay_eq}, one obtains that
\begin{equation}\label{int_s_z}
\begin{aligned}
	\int_{\Gamma_+^\epsilon} \det(\mathcal{S}^z)^{-1}\frac{d}{dz}\det(\mathcal{S}^z)dz&=\int_{\Gamma_+^\epsilon}\det(M_-^{1/z})^{-1}\frac{d}{dz}\det(M_-^{1/z}) - \det(M_-^z)^{-1}\frac{d}{dz}\det(M_-^z)dz\\&=-\left(\int_{\Gamma_+^\epsilon}+\int_{\Gamma_-^\epsilon}\right)\det(M_-^z)^{-1}\frac{d}{dz}\det(M_-^z)dz\\&=-\lim\limits_{\delta \to 0}\left(\int_{\Omega_+^{\epsilon,\delta}}+\int_{\Omega_-^{\epsilon,\delta}}\right)\det(M_-^z)^{-1}\frac{d}{dz}\det(M_-^z)dz.
\end{aligned}	
\end{equation}
Using \eqref{J_b}, \eqref{J_h_+}, \eqref{J_h_-} and \eqref{int_s_z} one arrives at 
\begin{equation}
	\begin{aligned}
		2\pi \imath (J_b+\frac{1}{2}J_h-L)&=\lim\limits_{\epsilon\to 0}\lim\limits_{\delta \to 0}\left(\int_{\Omega_{\epsilon,\delta}}-\int_{l_+^{\epsilon,\delta}}-\int_{l_-^{\epsilon,\delta}}\right)\det(M_-^z)^{-1}\frac{d}{dz}\det(M_-^z)dz\\&=\lim\limits_{\epsilon\to 0}\lim\limits_{\delta \to 0} \left(\int_{\Omega_+^{\epsilon,\delta}}+\int_{\Omega_-^{\epsilon,\delta}}\right)\det(M_-^z)^{-1}\frac{d}{dz}\det(M_-^z)dz\\&=-\lim\limits_{\epsilon \to 0}\int_{\Gamma_+^\epsilon}\det(\mathcal{S}^z)^{-1}\frac{d}{dz}\det(\mathcal{S}^z)dz,
\end{aligned}
\end{equation}
where $J_h=J_h^++J_h^-$.
 \hfill $\Box$ 

\vspace{.2cm}

\noindent {\bf Acknowledgements:}
This research was supported by CONACYT, FORDECYT-PRONACES  429825/2020 (proyecto apoyado por el FORDECYT-PRONACES, PRONACES/429825), recently renamed project CF-2019 / 429825. Further support came from the project PAPIIT-DGAPA-UNAM IN101621. M. B. is a Fellow of the Sistema Nacional de Investigadores (SNI).  The work  of H. S.-B. was also partially funded  by the grant DFG SCHU 1358/6-2. 

\appendix

\section{Appendix}

This appendix recollects a  some technical statements that are used in the main text.

\begin{theo}[Volterra equation, {Lemma}~7.8 in \cite{Tes}, and Theorem~26 in \cite{BFGS}]\label{volterra} 
	{Let  $g \in \ell^{\infty}(\mathbb{N},\Mm)$ and $K(n,m) \in \Mm$ for each $m,n\in \mathbb{N}$. 
		Consider the Volterra equation}
	\begin{align} \label{volt}
	f(n)=g(n)+\sum_{m=n+1}^{\infty}K(n,m)f(m),
	\end{align}
	and suppose there is a sequence $M\in \ell^{1}(\mathbb{N},\mathbb{R})$  such that $\|K(n,m)\|\leq M(m)$ for all $m, n\in \mathbb{N}$. Then Eq. \eqref{volt} has a unique solution $f\in \ell^\infty(\mathbb{N},\Mm)$. Moreover, if $g(n)$ and $K(n,m)$ depend  continuously (resp. holomorphically) on a parameter $z$ (for every $n$),  $M$ does not depend on $z$, and $g(n)$ is uniformly bounded w.r.t. $n$ and $z$, then the same is true for $f(n)$.
\end{theo}
\begin{lemma}[Gronwall lemma]\label{gronwall}
	Let $\alpha$ a real positive number and $(w_n)_{n\in \mathbb{N}}, (u_n)_{n\in \mathbb{N}}$ real positive sequences such that 
	$$\sum_{j=1}^\infty w_j <\infty, \qquad u_n\leq K , \ n\in \mathbb{N},$$ for some $K\in \mathbb{R}$ and 
	\begin{equation}\label{hip}
		u_n \leq \alpha + \sum_{j=n+1}^\infty w_j u_j. 
	\end{equation} 
	Then for all $n \in \mathbb{N}$, it follows that
	$$u_n \leq \alpha \exp\left(\sum_{j=n+1}^{\infty}w_j\right).$$
\end{lemma}
 \noindent {\bf Proof.} 
Let us provide a proof as this was already stated without proof in \cite{BFGS}. Let us define the functions $W,U : \RM \to [0, \infty)$ by setting
$$
U|_{[-n ,  - n +1)} = u_n ,  \qquad W|_{[-n,- n + 1)} = w_n, \qquad n\in\NM,
$$
and both $U$ and $W$ vanish on $ [0, \infty) $. For every $t \in [-n, -n+1)$, one has that 
\begin{align}\label{stupi}
U(t) = u_n \leq \alpha + \sum_{j = n+1}^\infty w_ju_j = \alpha + \int_{-\infty}^{-n } W U \leq \alpha +  \int_{-\infty}^{t } WU.  
\end{align}
For the rest of the proof, one argues as in the proof of the Gronwall lemma for the continuous case. We provide a few lines with the key steps, for the convenience of the reader. Let us define $V(t) = e^{- \int_{-\infty}^t W }\int
_{-\infty}^t WU$. It is  clear that $\frac{d}{dt} V(t) = e^{- \int_{-\infty}^t W } W(t)[ U (t) -  \int
_{-\infty}^t WU    ] \leq \alpha  e^{- \int_{-\infty}^t W } W(t), $ for every $  t \notin -\mathbb{N} \cup \{ 0 \} $. Integrating, one gets 
$$
V(t) \leq \int_{-\infty}^t  \alpha  e^{- \int_{-\infty}^s W } W(s)  = \alpha ( 1 - e^{- \int_{-\infty}^t W } ).  
$$
This implies that
$$
\int_{-\infty}^t WU \leq \alpha e^{ \int_{-\infty}^t W } - \alpha, 
$$
which together with \eqref{stupi} implies $ u_n  =   U(-n) \leq  \alpha e^{ \int_{-\infty}^{-n} W } =
 \alpha e^{  \sum_{j = n+1}^{\infty} w_j }  $. 
 \hfill $\Box$

\begin{lemma}\label{volteo}
	Let $h: \mathbb{S}^1\setminus \{-1,1 \} \to \mathbb{C}$ a continuously differentiable function and $ \Gamma_+$ a  curve on $ \mathbb{S}^1 \setminus \{-1,1 \}  $ that is  parameterized by a differentiable function $\gamma_+ : [0,1] \to 
\mathbb{S}^1\setminus \{-1,1 \}	
	$. We assume  that $0 \notin h(\Gamma_+)$. Let $r:\mathbb{C}\setminus\{0\} \to \mathbb{C}$ the function $r(z)=1/z$ and $\Gamma_-:=r(\Gamma_+)$ parameterized by $\gamma_-(t):=1/\gamma_+(1-t)$. The following identity holds true: 
	$$\int_{\Gamma_-}  \frac{h'}{h} = -\int_{\Gamma_+} \frac{(h\circ r)'}{h\circ r}. $$    
\end{lemma}
 \noindent {\bf Proof.} 
	\begin{equation*}
	\begin{aligned}
\int_{\gamma_-} \frac{h'}{h} &=\int_0^1 \frac{h'\circ \gamma_-(t)}{h\circ \gamma_-(t)} (\gamma_-)'(t)dt =-\int_1^0 \frac{h'\circ \gamma_-(1-t)}{h\circ \gamma_-(1-t)} (\gamma_-)'(1-t)dt\\&=\int_0^1 \frac{h'(1/\gamma_+(t))}{h ( 1/\gamma_+ (t))}1/(\gamma_+(t))^2 (\gamma_+)'(t)dt=\int_{0}^{1} \frac{h'\circ r(\gamma_+(t))}{h \circ r (\gamma_+ (t))}1/(\gamma_+(t))^2 (\gamma_+)'(t)dt\\&=-\int_0^1 \frac{(h\circ r)'(\gamma_+(t))}{h\circ r(\gamma_+(t))} (\gamma_+)'(t)dt=-\int_{\Gamma_+} \frac{(h\circ r)'}{h\circ r},
	\end{aligned}		
	\end{equation*}
	where the penultimate equality follows from $(h\circ r)'(z)=- (1/z^2) h'\circ r(z)$.
 \hfill $\Box$ 
\begin{lemma}\label{aux}
	Let $r:\overline \DM \to \mathbb{C}$ a continuous function (or continuous in a closed neighborhood of 1 in $ \overline \DM   $) such that it is analytic on $\DM $ (or analytic in the intersection of a neighborhood of 1 with $\DM$) and $r(1)=c\neq 0$. Let $(\gamma_n)_{n\in \mathbb{N}}$ a sequence of curves satisfying  
	$$\gamma_n \subset D_n\cap \DM,$$ where $D_n=\{z\in \DM:|z-1|<1/n\}$.
	 It follows that
	$$\lim_{n\to \infty} \int_{\gamma_n}\frac{r'(z)}{r(z)}dz=0.$$ 
\end{lemma}
 \noindent {\bf Proof.} 
	Since $c\neq 0$, we can assume w.l.o.g. that $\Re(c)>0$ (we multiply everything by a constant complex number).  Let $B$  an open set such that $c\in B$ and  
$$(-\infty, 0]\cap B = \emptyset.$$ 
We set $$\log: \mathbb{C}\setminus[-\infty,0] \to \mathbb{C}$$ an analytic branch of logarithm.  By the continuity of $r$, there exist $n\in \mathbb{N}$ such that $r(\overline{D_n\cap \DM}) \subset B$.  Then, for $m\geq n$,  the function $$\log\circ r : \overline{D_m\cap \DM} \to \mathbb{C}$$  is continuous, and analytic on $D_m\cap \DM$. For $m\geq n$, we calculate
	$$\int_{\gamma_m} \frac{r'(z)}{r(z)} dz = \int_{\gamma_m} (\log\circ r)'= \log\circ r(\gamma_m(1)) -\log\circ r(\gamma_m(0)).$$  
	The desired result follows from the fact that  $\log \circ r$ is  continuous on $ \overline{D_m\cap \DM}$ and, consequently, $$\gamma_m(1)-\gamma_m(0) \to 0, \ m \to \infty .$$   
 \hfill $\Box$ 

\begin{proposi}[Schur formula for the determinant]\label{schur}
	Let $M=\begin{pmatrix}
		A & B\\
		C & D
	\end{pmatrix}$ be a block matrix with square matrices $A$ and  $D$. If $D$ is invertible then 
$$\det(M) = \det(D)\det(A-BD^{-1}C).$$
\end{proposi}


\begin{thebibliography}{99}


\bibitem{Agranovich} Z.~S.~Agranovich, V.~A.~Marchenko, {\sl The inverse problem of scattering theory}, (Courier Dover Publications, 2020).

\bibitem{ACP} T.~Aktosun, A.~E.~Choque-Rivero, V.~G.~Papanicolaou, {\sl On the bound states of the discrete Schr\"odinger equation with compactly supported potentials}, Electron. J. Differential Equations 2019, Paper No. 23 (2019).

\bibitem{AW0} T. Aktosun, R. Weder,  {\sl High-energy analysis and Levinson's theorem for the selfadjoint matrix Schr\"odinger operator on the half line}, J. Math. Phys.  {\bf 54}, 012108 (2013).

\bibitem{AW} T.~Aktosun, R.~Weder, {\sl Direct and Inverse Scattering for the Matrix Schr\"odinger Equation}, (Springer International, Switzerland, 2020). 

\bibitem{AN} A.~I.~Aptekarev, E.~M.~Nikishin, {\sl The scattering problem for a discrete Sturm-Liouville operator,} Math. USSR Sbornik {\bf 49}, 325-355 (1984).

\bibitem{BAC1} E.~Bairamov, Y.~Aygar, S.~Cebesoy, {\sl Spectral analysis of a selfadjoint matrix-valued
discrete operator on the whole axis} J. Nonlinear Sci. Appl. {\bf 9}, 4257-4262 (2016).

\bibitem{BAC2} E.~Bairamov, Y.~Aygar, S.~Cebesoy, {\sl  Investigation of Spectrum and Scattering Function of Impulsive Matrix Difference Operators}, Filomat {\bf 33:5}, 1301-1312  (2019).

\bibitem{BFS} M.~Ballesteros, G.~Franco~C\'ordova, H.~Schulz-Baldes, {\sl Analyticity properties of the scattering matrix for matrix Schr\" odinger operators on the discrete line}, J. Math. Anal. Appl. {\bf 497}, 124856 (2021).

\bibitem{BFGS} M.~Ballesteros, G.~Franco, G.~Garro, H.~Schulz-Baldes, {\sl Band edge limit of the scattering matrix for quasi-one-dimensional discrete Schr\"odinger operators}, Complex Analysis and Operator Theory  {\bf 16}, 1-31 (2022).

\bibitem{BS} J.~Bellissard, H.~Schulz-Baldes, {\sl Scattering theory for lattice operators in dimension $d\geq 3$},  Rev. Math. Phys. {\bf 24}, 1250020 (2012).

\bibitem {Cas} K.~M.~Case, M.~Kac, {\sl A discrete version of the inverse scattering problem}, J. Math. Phys. {\bf 14}, 594-603 (1973).

\bibitem{CS} A.~M.~Childs, D.~J.~Strouse, {\sl Levinson's theorem for graphs}, J. Math. Phys. {\bf 52}, 082102 (2011).

\bibitem{Dei} P.~Deift, E.~Trubowitz, {\sl Inverse scattering on the line}, Comm. Pure Appl. Math. {\bf 32}, 121-251 (1979).

\bibitem{EMT} I.~Egorova, J.~Michor, G.~Teschl, {\sl Scattering theory for Jacobi operators with quasi-periodic background}, Commun. Math. Phys. {\bf 264}, 811-842 (2006).

\bibitem{Gus} G.~Sh.~Guseinov, {\sl Determination of an infinite Jacobi matrix from scattering data}, Dokl. Akad. Nauk SSSR {\bf 227}, 1289-1292 (1976).

\bibitem{Gus2} G.~Sh.~Guseinov, {\sl The inverse problem of scattering theory for a second order difference equation on the whole real line}, (Russian) Dokl. Akad. Nauk SSSR {\bf 230}, 1045-1048 (1976).

\bibitem{Gus3} G.~Sh.~Guseinov, {\sl The scattering problem for an infinite Jacobi matrix}, (Russian) Izv. Akad. Nauk Armyan. SSR Ser. Mat. {\bf 12}, 365-379 (1977).

\bibitem{HKS} D.~B.~Hinton, M.~Klaus, J.~K.~Shaw, {\sl Half-bound states and Levinson's theorem for discrete systems},  SIAM J. Math. Analysis {\bf 22}, 754-768 (1991).

\bibitem{IT} H.~Inoue, N.~Tsuzu, {\sl Schr\"odinger Wave Operators
on the Discrete Half-Line}, Integr. Equ. Oper. Theory {\bf 91}, 1-12 (2019).

\bibitem{KR} J. Kellendonk, S. Richard, {\sl The topological meaning of Levinson's theorem, half-bound states included}, 
{J. Phys. A: Math. Theo. {\bf 41}, 295207-295217 (2008)}.

\bibitem{Mar} L.~Mart\'inez Alonso, E.~Olmedilla, {\sl Trace identities in the inverse scattering transform method associated with matrix Schr\"odinger operators}, J. Math. Phys. {\bf 23}, 2116-2121 (1982).

\bibitem{NRT} H.~S.~Nguyen, S.~Richard, R.~Tiedra de Aldecoa, {\sl Discrete Laplacian in a half-space with a periodic surface
potential I: Resolvent expansions, scattering matrix, and wave operators},  Math. Nachr. {\bf 295}, 912-949 (2022).

\bibitem{Ser}   V.~P.~Serebryakov, {\sl The inverse problem of scattering theory for difference equations with matrix coefficients}, Doklady Akad. Nauk  {\bf 250},  562-565 (1980). 

\bibitem{Tes}  G.~Teschl, {\sl Jacobi operators and completely integrable nonlinear lattices},  (AMS, Providence, 2000).









\end{thebibliography}
\end{document}